# Impact of the scattering physics on the power factor of complex thermoelectric materials

Patrizio Graziosi[1,2*], Chathurangi Kumarasinghe[1], Neophytos Neophytou[1]

[1] School of Engineering, University of Warwick, Coventry, CV4 7AL, UK
[2] CNR – ISMN, v. Gobetti 101, 40129, Bologna, Italy
[*] Patrizio.Graziosi@warwick.ac.uk

## Abstract

We assess the impact of the scattering physics assumptions on the thermoelectric properties of five Co-based p-type half-Heusler alloys by considering full energy-dependent scattering times, versus the commonly employed constant scattering time. For this, we employ DFT bandstructures and a full numerical scheme that uses Fermi's Golden Rule to extract the momentum relaxation times of each state at every energy, momentum, and band. We consider electron-phonon scattering (acoustic and optical), as well as ionized impurity scattering, and evaluate the qualitative and quantitative differences in the power factors of the materials compared to the case where the constant scattering time is employed. We show that the thermoelectric power factors extracted from the two different methods differ in terms of: i) their ranking between materials, ii) the carrier density where the peak power factor appears, and iii) their trends with temperature. We further show that the constant relaxation time approximation smoothens out the richness in the bandstructure features, thus limiting the possibilities of exploring this richness for material design and optimization. These details are more properly captured under full energy/momentum-dependent scattering time considerations. Finally, by mapping the conductivities extracted within the two schemes, we provide appropriate density-dependent constant relaxation times that could be employed as a fast first–order approximation for extracting charge transport properties in the half-Heuslers we consider.

## 1. Introduction

Thermoelectric generators (TEG) convert heat flow into useful electrical power and could provide energy savings and reduced dependence on fossil fuels. TEG are based on thermoelectric (TE) materials, whose ability to convert heat into electricity is quantified by the dimensionless figure of merit $ZT = \sigma S^2 T/\kappa$, where $\sigma$ is the electrical conductivity, $S$ is the Seebeck coefficient, and $\kappa$ is the thermal conductivity. Some of the best bulk TE materials have $ZT \sim 1$, which is, however, insufficient for large scale implementation. It is estimated that a $ZT > 3$ based on inexpensive, non-toxic, and abundant materials will increase the applications for thermoelectricity by 10-fold and will empower large scale commercialization.[1] At present, the efforts to increase $ZT$ mostly focus on the reduction of the lattice thermal conductivity by acting on the granular structure of the samples, a strategy that resulted in $ZT$ values above 2,[1-4] but strategies to increase the power factor $\sigma S^2$ (PF) can be beneficial as well.[5]

New classes of materials started to emerge in the last several years, which could bring large improvements in the PF. Half-Heusler alloys, SnSe, PbTe, and BiTe based compounds, clathrates, skutterudites, to name a few, have complex electronic bandstructures with multiple anisotropic bands in multiple valleys placed close to the conduction and valence band edges, which are thought to be beneficial to the PF.[4,6] Half-Heuslers alloys, in particular, combine thermal and mechanical stability, low toxicity, reasonable price, and good TE performance in terms of high power factors.[7-10]

Theoretical studies to assess the performance of such materials are usually based on extracting the bandstructure using density-functional-theory (DFT), and then calculate the TE coefficients within the semi-classical Boltzmann Transport Equation (BTE).[11] However, due to the complexity of the bandstructures and scattering physics, the constant relaxation time approximation is usually employed within the BTE. Well-established publicly available software are also available towards this effort, each having different capabilities and strengths, namely BoltzTraP,[12] BoltzWann,[13] aMoBT[14] and LanTraP.[15] In reality, however, the scattering rates are energy, momentum, and band dependent, and multiple scattering mechanisms such as electron-phonon scattering, ionized impurity scattering, alloy scattering, boundary scattering, are contributing to the scattering times, each having a distinct energy/momentum dependence (elastic or inelastic, isotropic or anisotropic).[16,17] In the light of numerous studies undertaken recently towards large data materials screening and ranking, not only for TE materials, but for other applications as well,[2,3] it is imperative that some of these details are considered to extract the electronic properties, or at least the consequences of omitting them, understood and quantified.[18]

In this work, we relax the constant relaxation time approximation with a code that can consider the full energy, momentum, and band dependence of the scattering rates, considering carrier scattering with phonons and ionized impurities. We study five p-type Co-based half-Heusler alloys, whose complex valence bands[7, 19] make them an excellent tool to assess the impact of the scattering physics in the transport properties of complex materials: TiCoSb, ZrCoSb, HfCoSb, ZrCoBi and NbCoSn. (Note that in the present context, 'complex' is commonly used to refer to the rich features, far away the simple parabolic shapes, with no reference to 'imaginary' components). We consider a full 3D bandstructure as extracted from DFT and scattering rates due to acoustic and optical phonons, as well as ionized impurity scattering, extending our previous 2D and 1D implementations in other mateirals.[20-23] We demonstrate that different qualitative findings are reached, with regards to materials PF rankings, optimal carrier density and temperature trends, when comparing energy/momentum dependent versus constant scattering times. The paper is organized as follows: in Section 2 we describe the theoretical and computational methodology; in Section 3 we present and discuss our results in terms of the impact of the scattering physics on the charge transport properties of the Heusler materials and their power factor, and finally, in Section 4 we conclude.

## 2. Approach

The approach consists of three stages: i) calculation of the bandstructures using DFT, ii) numerical extraction of the scattering rates, and iii) use of the BTE for the calculation of the TE coefficients. The electronic bands are calculated within the DFT scheme using the Quantum Espresso package.[18, 24, 25] Projector augmented wave technique was used with the PBE-GGA functional and a kinetic energy cut-off greater than 60 Ry was used for the wave functions. An energy convergence criterion of $10^{-8}$ Ry for self-consistency was adopted throughout our calculations. The 3D bandstructure was calculated using a 51x51x51 Monkhorst–Pack $k$-point mesh on the primitive unit cell of the reciprocal lattice. The $k$-points coordinates, originally described in the coordinate system of the reciprocal unit cell, are expressed in orthogonal coordinates to work in a cartesian system and we then calculate the transport quantities as x,y,z tensors.

The bandstructures for the five half-Heuslers under consideration are shown in **Fig. 1**. There, we only show the 1 eV energy range into the valence band (VB) and the conduction band (CB) edges, which participate in transport (we only consider hole transport in this work). Multiple bands of different curvatures in different directions compose the bandstructure and participate in transport. **Figure 1f** shows an example of the iso-surfaces in 3D at $E$ = -0.1 eV to illustrate the complexity of the bands in 3D. In the transport calculations we consider the lower VB energies up to -0.7 eV below the valence band edge as indicated by the shaded regions, that is around $7k_BT$ beyond the highest Fermi level for the $T$ = 900 K, which is the maximum temperature we use. **Figure 1g** shows the primitive cell of the TiCoSb as an example, while the conventional zincblende unit cell is shown in **Fig. 1h**. These half-Heusler materials are in fact compounds (the term alloy endures for historical reasons).

As an illustration of the numerical complexity of the computation, in **Fig. 2a** we show an iso-energy surface for TiCoSb at $E$ = -0.12 eV into the valence band. The surface has elongated tubes and flat regions, an anisotropy that is thought to be very beneficial for the Seebeck coefficient.[3, 6] For the numerical calculation of the scattering rates and the transport state properties needed in the BTE (velocities and density-of-states (DOS)), first we transform information with respect to the $E(\boldsymbol{k})$, obtained from DFT, into $\boldsymbol{k}(E)$. Thus, for every energy we gather the information of all $\boldsymbol{k}$-states with their velocities and their DOS.[20-22] We show the sampling of the same iso-energy surface in **Fig. 2b**. The surface contains tens of thousands of points, to each of them a velocity vector and density-of-states is assigned. Each $\boldsymbol{k}$-point is considered as an initial state $\boldsymbol{k}$ for the carrier that can scatter in all final states $\boldsymbol{k'}$. As the selection rules and details of the strength of the electron-phonon coupling of each initial state to all other states individually is not yet well established for half-Heulser alloys in general, in this work we consider both intra-valley and inter-valley scattering events (using deformation potential theory). The final states can reside on the same surface in the case of intra-band scattering, or on the surface of different bands in the case of inter-band scattering, as sketched in **Fig. 2b**, and **2c** (orange dots indicate states from a different band compared to the initial one), respectively. In the case of inelastic scattering, the initial states are scattered into states with final energies $E_i \pm \hbar\omega$ (red/green dots in **Fig. 2d**).

In addition, for each possible pair ($\boldsymbol{k}$, $\boldsymbol{k'}$) we have a scattering rate for each relevant scattering mechanism. Thus, we separately include all energetics from the contributions of different scattering mechanisms, especially ionized impurities that play a major role at the high doping concentration of common TE materials. Yet, for the extraction of the scattering rates, we require deformation potentials, sound velocities, phonon energies, and dielectric constants, parameters that are still not well established for Heuslers or most of the new generation advanced TE materials. We adopt these parameters from the literature[26] and the Materials Project database.[27] Table I reports the used parameters and their values.

We consider elastic scattering with acoustic phonons (ADP), both intra- and inter-band, inelastic scattering with optical phonons (ODP), being both intra- and inter-band, and scattering with ionized impurities (IIS), considered as only intra-band. The polar optical phonon (POP) scattering is a relevant scattering mechanism in compound semiconductors, and is mainly active for carriers residing in valleys centred around $\Gamma$.[28] However, due to the lack of the necessary relevant parameters in the literature and the lack of knowledge about the selection rules for these compounds, we do not consider POP here. This could overestimate our conductivity calculations at some degree, however, we do not expect any qualitative changes. The overall momentum relaxation time for each scattering mechanism *(i)* of a carrier in a state (***k***,*n*,*E*) is derived from the scattering rate between the considered state and all the possible final states as defined by intra- and inter-band considerations, energy and momentum conservation, by:

$$\frac{1}{\tau_{x,k,n,E}^{(i)}} = \frac{1}{(2\pi)^3} \sum_{k'} |S_{k,k'}^{(i)}| \left(1 - \frac{v_{x,k'}}{v_{x,k,n,E}}\right) \tag{1}$$

The sum in Eq. (1) runs on all the possible final states ***k'*** and $|S_{k,k'}|$ is the transition rate. The $\left(1 - \frac{v_{k'}}{v_k}\right)$ term is an approximation for momentum relaxation, but uses the state velocities instead of the momenta, which is a generalization of the case where multiple bands with multiple effective masses at different regions of the Brillouin zone participate in transport.[21, 22, 29, 30] Since these half-Heuslers are non-ferromagnetic the carriers only scatter in states of the same spin as in common semiconductors.[28, 31]

$|S_{k,k'}|$ is derived from the Fermi's Golden Rule for different scattering mechanisms, (acoustic deformation potential, optical deformation potential, and ionized impurity scattering), in the usual way, as:[30, 31]

$$\left|S_{k,k'}^{(\text{ADP})}\right| = \frac{1}{V_c} \frac{\pi}{\hbar} D_{\text{ADP}}^2 \frac{k_B T}{\rho v_S^2} \, \delta(E_{k'} - E_k) \tag{2a}$$

$$\left|S_{k,k'}^{(\text{ODP})}\right| = \frac{1}{V_c} \frac{\pi D_{\text{ODP}}^2}{\rho \omega} \left(N_{\omega,\text{BE}} + \frac{1}{2} \mp \frac{1}{2}\right) \, \delta(E_{k'} - E_k \pm \hbar\omega) \tag{2b}$$

$$\left|S_{k,k'}^{(\text{IIS})}\right| = \frac{1}{V_c} \frac{2\pi}{\hbar} \frac{Z^2 q_0^4}{k_S^2 \varepsilon_0^2} \frac{N_{\text{imp}}}{\left(|k-k'|^2 + \frac{1}{L_D^2}\right)^2} \, \delta(E_{k'} - E_k). \tag{2c}$$

Above, $V_C$ is the crystal volume, $D_{ADP}$ is the acoustic deformation potential, $\rho$ is the mass density, $v_s$ is the sound velocity computed as $v_s = \frac{1}{3}s_l + \frac{2}{3}s_t$, which is important when the bands are not isotropic,[16] where $s_l = \sqrt{\frac{K_V + 4/3 G_V}{\rho}}$ and $s_t = \sqrt{\frac{G_V}{\rho}}$ are the longitudinal and transverse sound speeds and $K_V$ and $G_V$ are the bulk and shear modulus. $D_{ODP}$ is the optical deformation potential, $\omega$ is the longitudinal optical phonon frequency in the single mode approximation with constant frequency over the entire reciprocal lattice unit cell and $N_{\omega,BE}$ is its population density given by the Bose-Einstein statistics where the '+' and '−' signs indicate the absorption and emission processes, respectively. $N_{imp}$ is the impurity density, $Z$ the impurity charge, $\varepsilon_0$ and $k_s$ are the vacuum and the static relative permittivities. $L_D$ the Debye screening length in 3D defined as:

$$L_D = \sqrt{\frac{k_s \varepsilon_0}{q_0} \frac{\partial E_F}{\partial n}} \tag{3}$$

where $n$ is the carrier density and $\partial n/\partial E_F$ is the variation of the carrier density with respect to the Fermi level, which is temperature and doping dependent.[30, 31] The explicit use of $\partial n/\partial E_F$ enables us to apply the equation also in the degenerate doping conditions. In the calculations, the doping concentration is assumed to be equal to the carrier density at a specific Fermi level position, which is an input to the code, as calculated at 300 K. We consider the Fermi level movement upon temperature by allowing it to shift in order to ensure keep constant carrier density. The wavefunctions overlap integral is approximated to the unity.

In Eq. (2) the delta function ensures the energy conservation while the momentum conservation is assured by the phonon or crystal momentum. The delta functions define a constant energy surface and the evaluation of the scattering rate, Eq. (1), becomes numerically a surface integral.[32, 33] Computationally, the sum in Eq. (1) runs over the points of the constant energy surface. In this way a density of states for each individual final $k'$-state is computed as $g_{k',n,E} = \frac{dA_{k',n,E}}{\hbar |\vec{v}_{k',n,E}|}$, where $dA_{k',n,E}$ is the corresponding surface element (on the iso-energy surface at energy $E$ of band $n$) associated to each $k'$-state and $\vec{v}$ its band velocity.[20, 33] Using the regular discretization of the reciprocal unit cell that the DFT bandstructure is computed on, we extract the constant energy surfaces as a set of points in the $k$-space for which the individual $g_{k,n,E}$ is essentially replacing the delta function and the crystal volume terms. Details for the extraction of $dA_{k,n,E}$ are presented in the supplementary material. The deformation potentials are taken from the literature, where they have been computed from the electron-phonon matrix interaction within the EPW code.[26] Since these materials are compounds, no alloy scattering is considered.

Within the linearized BTE formalism, the charge transport TE coefficients are defined as:[20-22]

$$\sigma = q_0^2 \int_E \Xi(E) \left(-\frac{\partial f_0}{\partial E}\right) dE, \qquad (4a)$$

$$S = \frac{q_0 k_B}{\sigma} \int_E \Xi(E) \left(-\frac{\partial f_0}{\partial E}\right) \frac{E-E_F}{k_B T} dE, \qquad (4b)$$

where $E_F$, $T$, $q_0$, $k_B$, are the Fermi level, the absolute temperature, the electronic charge and the Boltzmann constant respectively, $f_0$ is the equilibrium Fermi distribution. $\Xi(E)$ is the so-called transport distribution function defined as:[13]

$$\Xi(E) = \frac{2}{(2\pi)^3} \frac{1}{V_c} \sum_{k,n}^{BZ} v_{k,n,E}^2 \tau_{k,n,E} \delta(E_k - E) = \frac{2}{(2\pi)^3} \sum_{k,n}^{\mathfrak{L}_E^n} v_{k,n,E}^2 \tau_{k,n,E} g_{k,n,E} \qquad (5)$$

where $v_{k,n,E}$ is the band velocity of the charge carrier in the state defined by the wavevector $k$ in the band $n$ at energy $E$, $\tau_{k,n,E}$ its relaxation time (as defined in Eq. (1)) and $g_{k,n,E}$ its density-of-states (DOS) associated with the individual states, BZ stands for the Brillouin Zone and $\mathfrak{L}_E^n$ represents the surface of constant energy $E$ for the band of index $n$.[21-23] The first sum in Eq. (5) runs over all the $k$-states and the bands of the BZ. The delta-function which picks up only the states at energy $E$, defines a surface of energy $E$ for each band. The second sum in Eq. (5) runs on all the points of these surfaces, for all bands, and returns an energy dependent quantity evaluated on all the iso-energy surfaces $\mathfrak{L}_E^n$. The triad ($k,n,E$) defines uniquely each transport state that, as above, has DOS $g_{k,n,E} = \frac{dA_{k,n,E}}{\hbar |\vec{v}_{k,n,E}|}$, where $dA_{k,E,n}$ is the surface area element associated to each $k$ state in the band $n$ at energy $E$.[20, 33] Each individual ($k,n,E$) state has its own specific $\tau_{k,n,E}$ that is composed from the scattering time of each mechanism (Eq. (1)), all combined together using Matthiessen's rule.

The surface element $dA_{k,E,n}$ is in general complicated to be extracted for each state on a surface of an arbitrary shape. In general, this can be achieved by Delaunay Triangulation of the energy surfaces, which, however, we found computationally expensive, given that the energy surfaces in these materials extend in the entire Brillouin zone. However, once we have the $k$-points that reside on a constant energy surface, we find that the area of the circle having radius equal to half the average distance between the specific ($k,n,E$) point and its nearest neighbours on the iso-energy surface, is a very good approximation for its surface element. For this, the nearest neighbours are defined as the nearby points within $\sqrt{2} \cdot dk$ where $dk$ is the average distance between neighbour points in the initial input regular $k$-mesh from DFT. We have validated this approach versus parabolic and non-parabolic band cases that have analytical forms for the DOS with excellent agreement, as well as for a Heusler material in comparison to the full Delaunay triangulation method. The method and its validation are detailed in the supplementary material. The scattering treatment (Eq. 2-4) has also been validated for the case of silicon.

### 3. Results and Discussion

We start our investigation by comparing the PF of the five materials under consideration under three different scattering scenarios at 300 K. **Figure 3** shows the power factor (PF) of the five Heuslers versus the relative position of the Fermi level $\eta_F$, that essentially corresponds to the density or doping level, for the cases of: i) constant relaxation time approximation, $\tau_c$, ii) energy dependent phonon-limited scattering, $\tau_{ph}(E)$, and iii) energy dependent phonon plus ionized impurity scattering, $\tau_{ph,IIS}(E)$. When we consider the scattering processes we deal with a state dependent relaxation time $\tau(k,n,E)$, as in Eq. (5), but for simplicity below we use the notation $\tau(E)$. Note that the valence band edge is set to zero and a negative $\eta_F$ means that the Fermi level is pushed into the bands – i.e. degenerate conditions. In all sub-figures orange lines are for HfCoSb, blue for TiCoSb, purple for ZrCoSb, red for NbCoSn, and black for ZrCoBi. In the constant relaxation time approximation (**Fig. 3a**) we arbitrarily chose $\tau_c$ = 10 fs as it's a typical used value in the thermoelectric literature. This value is commonly employed for computational studies related to thermoelectric materials.[34] It is in the neighbourhood of values estimated in experimental settings (usually for polycrystalline, non-defect free materials). It arises from the general lack of data about single crystal mobility in half-Heusler alloys, across temperatures, and across doping values. It is understood that more refined values could be used to match more sophisticated calculations, but any other number will only have a quantitative effect on our results. In this work, however, we are focusing on the qualitative trends that the energy dependence of scattering times brings, thus, we still employ the common to the thermoelectric literature $\tau_c$ = 10 fs.

This choice does not qualitatively affect the following discussion and materials ranking considerations. Under a constant relaxation time approximation, TiCoSb, NbCoSn and HfCoSb have the best performance while ZrCoSb has the worst performance.

We compare this scenario with the case of phonon-limited scattering (**Fig. 3b**), and then in the case of phonons plus impurity scattering limited PFs (**Fig. 3c**). We will not be considering any quantitative differences, as those will depend on the arbitrary chosen constant relaxation time, however, two important qualitative differences can be noted:

(i) The materials ranking can be different in the three scenarios. Under the constant relaxation time approximation $\tau_c$, TiCoSb, NbCoSn and HfCoSb have a similar performance, topping the PF values. In the energy dependent scattering cases, still NbCoSn and secondly ZrCoBi and HfCoSb hold the highest power factors, however the TiCoSb ranking drops to the lowest, even though it is one of the best performers in the constant relaxation time case. Moreover, ZrCoBi (black line) ranks second to last in the constant relaxation time scenario, while it ranks second/third when energy-dependent scattering rates are considered and looks more promising. Indeed, recent experimental data indicate that TiCoSb performs less than ZrCoBi in terms of both PF and ZT and indicate ZrCoBi as one of the best p-type half-Heusler alloys.[35]

(ii) The PF peaks appear when the $E_F$ is pushed into the bands at 0.1 eV in the constant scattering time case, whereas the peaks shift to lower Fermi energy position (and consequently doping levels) at $\eta_F = 0$, in the energy-dependent scattering cases.

Two experimental points from the literature are shown by the hexagons in **Fig. 3c**, measured at 300 K for ZrCoBi (black hexagon) and ZrCoSn (purple hexagon). To plot these points we have extracted the Fermi level position that corresponds to the measured carrier concentration.[35] The measured data indicate lower PFs since the materials are almost certainly polycrystalline with grain boundaries that are additional scattering centres,[36-38] whereas our simulations are for single crystals. In addition, solid solution effects may rise from heavy substitutional doping and affect the comparison, as solid solutions can have different elastic deformation potentials and dielectric constants, together with a possible alloy scattering contribution,[39] while the calculations consider single crystals cases. Note that the quantitative accuracy is also very sensitive to the accuracy of the input parameters (deformation potentials, dielectric constants, etc.), the possible role of polar phonons, the microstructure details (grain boundaries, nanoinclusions and defects). Despite all these unknowns, however, the agreement, for such a complex structure is within a factor of ~ 2 which provides credit to our energy-dependent computations.

To provide more indications about the qualitative and quantitative differences that appear because of considering different scattering specifics, and to investigate what makes the best PF performers, we analyse in detail the charge transport properties of NbCoSn and TiCoSb. The former performs very well in all the scattering scenarios while the order of the latter changes significantly in ranking. The electrical conductivity and the Seebeck coefficient are plotted versus $\eta_F$ at 300 K in **Figs. 4a** and **4b** for the energy dependent scattering time $\tau_{ph}(E)$ (solid lines) and for the constant relaxation time $\tau_c$ (dashed lines). With blue lines we show TiCoSb whereas with red lines the NbCoSn. Under the constant relaxation time approximation, TiCoSb and NbCoSn have very similar conductivity (dashed lines in **Fig. 4a**), but under the energy dependent scattering case the conductivity of NbCoSn is much higher than that of TiCoSb. The conductivity differences lead the shift in ranking, since their Seebeck coefficients are quite similar as seen on **Fig. 4b**.

The better performance of NbCoSn in the $\tau(E)$ case can be explained when we consider the shape of the transport distribution function $\Xi(E)$ in both cases. In the $\tau_c$ case, we have at first order $\Xi(E) \sim v^2(E)\,g(E)$, whereas in the $\tau(E)$ case, since at first order the relaxation time is proportional to the inverse of the DOS, we have $\Xi(E) \sim v^2$. Thus, in the former case, the DOS has a relevant contribution in determining the $\Xi(E)$, together with the velocity squared. In the $\tau(E)$ case, on the other hand, the $\Xi(E)$ function is at first order determined only by the velocity squared, rather than the DOS. In **Fig. 4c** we compare the DOS of the two materials. TiCoSb has a higher DOS in the energy region within 0.2 eV from the band edge, where the states that participate in transport are located. Thus, the higher DOS will benefit TiCoSb under the $\tau_c$ approximation over NbCoSn. The $\Xi(E)$ functions for these cases are show in **Fig. 4d**. In the $\tau_c$ case, due to larger velocities, NbCoSn performs slightly better (dashed red line versus dashed-blue line), despite the larger DOS of the TiCoSb. The difference in the velocities, however, provides to NbCoSn a much higher transport distribution function $\Xi(E)$ in the energy-dependent case (solid lines in **Fig. 4d**). Note, however, as observed in Table I, that NbCoSn has a higher mass density and sound velocity, which result in a lower ADP scattering rate, and together with a higher dielectric constant that enables higher screening, it also experiences weaker IIS. Thus, NbCoSn experiences higher velocities, and weaker scattering in general as well, which results in an improved transport distribution function, conductivity, and power factors under energy/momentum dependent scattering conditions.

An important issue that is observed in the $\Xi(E)$ in **Fig. 4d**, on the other hand, is the shape of the $\Xi(E)$ functions under $\tau_c$ (dashed lines) and $\tau_{ph}(E)$ (solid lines). They differ substantially when the energy-dependent scattering time is considered, while they are very similar in the constant relaxation time approximation. They are also much smoother in the $\tau_c$ case, whereas they have much richer features under the $\tau_{ph}(E)$ case. The reason is that under $\tau_c$, $\Xi(E) \sim v^2(E)g(E)$, and since $g(E) \sim 1/|v(E)|$, the product of the two quantities at first order smears out a lot of the bandstructure features of the materials, that would have provided significant variations in their performance. On the other hand, when we consider the $\tau_{ph}(E)$, we have a relaxation time that at first order is related to $1/g(E)$. This results in $\Xi(E) \sim v^2$, which enhances differences in the band details between materials (square vs linear dependence on the velocity). Thus, the important point is that the proper energy-dependent treatment of the scattering mechanisms captures the richness of the bandstructures whereas constant relaxation time approximation smoothens the differences that those bandstructures have induced into the transport properties. We note here that the fact that the Seebeck coefficient, *S*, is also proportional to the energy derivative of the DOS (or the conductivity, and in the more general case the $\Xi(E)$), may further bring richer trends for the power factor, which are also smeared out under the constant relaxation time approximation.

Thus, a scattering time treatment that is sensitive to the charge carrier state specific momentum and energy dependent relaxation time is necessary to grasp the details of the transport distribution functions and provides different ranking outcomes between materials compared to the contestant relaxation time approximation. However, this comes with the limitations of the uncertainty in the deformation potentials, but still the energy-dependence by itself contains a richer and more complete description.

Now we analyse the temperature dependence of the PF under the $\tau_c$ and $\tau(E)$ scattering scenarios. We first study the dependence of the PF on the carrier density at different temperatures between 300 K and 900 K. We only report on HfCoSb, which is anyway representative of all the five compounds studied. We consider the Fermi level movement upon temperature by allowing it to shift down in order to keep the carrier density constant when the Fermi distribution broadens. This approach is valid only in the so-called extrinsic region,[40] where the carriers' density is constant upon temperature. The extension of the extrinsic region in the half-Heuslers thermoelectric alloys is not known, but the compounds we investigate have a bandgap higher than 1 eV as computed by DFT (and also considering that DFT tends to underestimate the band gaps)[41-43] thus it seems an acceptable approximation.

Under the constant relaxation time approximation, the temperature increase leads to a monotonic increase in the PF, as shown in **Fig. 5a**. The increase appears because at the same carrier density, the $\partial f_0 / \partial E$ function broadens with temperature, which forces the Fermi level to shift to lower energies to keep charge neutrality. The shift in the $E_F$ increases the Seebeck coefficient (at the same carrier density). The PF peak position with density moves slightly, but in all cases, corresponds to a Fermi level positioned ~0.1 eV into the band. However, as in practice the phonon scattering rate increases with temperature, at first order one can consider a linear decrease of the constant relaxation time for a fairer comparison (as in Eq. 2a). The inset of **Fig. 5a** shows the case where the constant relaxation time is linearly scaled with the temperature from the 300 K value as $\tau_C(T) = \tau_{C(300)} \frac{300}{T}$. In this case, the curves almost collapse on each other, with still a slight decrease in the PF peak as $T$ increases, indicating that the increase in the Seebeck coefficient due to the shift in $E_F$ is compensated by the decrease of $\tau$ with temperature.

Similarly, in the electron-phonon scattering scenario, $\tau_{ph}(E)$, **Fig. 5b**, the reduction in conductivity with temperature is compensated with the increase in Seebeck due to the $E_F$ shift, with still the effect of the temperature decreasing the PF slightly. The main difference, from the scaled constant time case above, however, is the largest shift in the PF peak towards higher carrier densities. At the PF peak, the Fermi level is placed near the band edge, at ~ 0 eV, in much less degenerate conditions compared to the $\tau_c$ case where it is at ~ 0.1 eV into the band. The reason behind the different optimal $E_F$ positioning, can be understood again from the shape of the transport distribution functions in **Fig. 5d**, which shows the $\Xi(E)$ for the $\tau_c$ (dashed line) and the $\tau_{ph}(E)$ (solid line) cases. Qualitatively, the faster raise of the $\Xi(E)$ in the $\tau_{ph}(E)$ case near the band edge ($E = 0$ eV) gives a higher Seebeck coefficient, which shifts the $E_F$ placement for the optimal PF towards those energies. On the other hand, in the $\tau_c$ case, the slower raise of the $\Xi(E)$ function around the band edge signals a lower Seebeck coefficient, which sets the $E_F$ at energies more into the band to reach the optimal PF (at -0.1 eV). Considering the shift in the $E_F$ with the broadening of the Fermi distribution as the temperature increases, in the case where the Fermi level is placed at 0 eV, the shift is larger, compared to the degenerate/metallic case where the $E_F$ is placed -0.1 eV into the bands, and thus the optimal power factor density shift is larger with temperature in the $\tau_{ph}(E)$ case. This shift in the peak PF signals to the different PF trends with temperature that are often encountered in experiments, as well. In the relatively lightly-doped semiconductor materials, the PF decreases with $T$ (left vertical dashed line in **Fig. 5b**), whereas at heavily-doped conditions and metals (right placed dashed vertical line in **Fig. 5b**), the PF slightly increases with temperature.

Similar behaviour is observed when the scattering due to the ionized impurities is considered, $\tau_{ph,IIS}(E)$ in **Fig. 5c**. The behaviour is surprisingly very similar to the $\tau_{ph}(E)$ case, but the peak drop is less pronounced while the best doping indications do not change. Overall, introducing IIS reduces the PF peak by 2-3 times. Another important observation is the significantly large PFs that the calculations predict even with IIS scattering. This indicates that once the materials are optimized, their performance would largely increase.

The comparison arguments above become clearer when the PFs are plotted versus temperature ($T$) in **Fig. 6**. We pick two carrier densities to create the plots in **Fig. 6**, the ones shown in **Fig. 5b** with the vertical-dashed lines (these values are approximated within 20%, since the input to our simulator is a regular array for the Fermi level, rather than the density). As the temperature is raised, the Fermi level is allowed to shift to keep the carrier density constant. At 300 K, it turns out that the position of the Fermi level with respect to the band edge is the same for all the materials as well. In **Fig. 6a** we plot the PF for the constant relaxation time approximation calculations for the commonly used 10 fs value and for the case where this value is scaled with the temperature, as we did above in the inset of **Fig. 5a**. In the former case, as the temperature increases, the Seebeck coefficient $S$ increases because of the Fermi level shift away from the bands and leads to PF increase. In the latter case, when the constant relaxation time is linearly scaled with T, following the acoustic phonon scattering case, the trend becomes more complex and after a small initial increase, the PF slightly decreases because the improvement in $S$ cannot compensate the decrease of $\tau_c$.

The situation changes when electron-phonon limited scattering is considered, and different trends are observed depending on the carrier density. In **Fig. 6b** we plot the PF versus temperature for the two carrier density levels indicated by the dashed lines in **Fig. 5b**. We have chosen these values to reside in the left and right regions of the PF peak (see **Fig. 3b** and **3c**), since the behaviour changes from one case to the other. Recall that at optimal PF conditions, the Fermi level is positioned at the band edge, and the increase in temperature shifts the Fermi level away from the band edge to keep the carrier density constant. At the lower densities we consider, the Fermi level is in the bandgap to begin with at 300 K. As the temperature increases, the Fermi level is shifted even further away, drifting further from the optimal conditions, and the PF decreases. On the contrary, when the Fermi level is into the bands initially (at 300 K to begin with), raising temperature shifts it closer to the optimal situation at the band edge, and the PF increases. Another important feature is that the material ranking changes in some cases with temperature and doping level, emphasizing the importance of energy/momentum dependent treatment of the scattering.

These energy-dependent considerations of the energy/momentum/band index of the scattering time requires quite demanding calculations. In the calculations we present, each data point takes approximately ~ 25 hrs on a single CPU (when the ionized impurities scattering is considered), as well as significant memory requirements, > 15 Gb. Simulations for elevated temperatures are even more intensive, as higher temperatures require higher energy windows with more and larger iso-energy surfaces. Attempts to extract temperature dependent $\tau_c$ based on overall carrier scattering with acoustic phonons alone have also been proposed, leaving out complexities such as separating elastic/inelastic intra/inter-band scattering, in attempt to reduce computational costs.[44] However, it could be instructive to materials scientists, experimentalists and theorists who perform material screening studies, employ such methods, or even to still employ a constant relaxation time approximation (constant in energy) as a first order approximation to estimate the power factors of materials. On the other hand, one first step towards more detailed scattering is at least to include the Fermi level (or density) dependence.

In **Fig. 7**, we present effective constant relaxation times for the 5 half-Heusler alloys, as a function of carrier density, Fermi level position $\eta_F$ in the Figure, at room temperature. We consider the two scattering situations when only electron-phonon scattering is considered, $\tau_{ph}(E)$ (**Fig. 7a**), and when the ionized impurity scattering is added as well, $\tau_{ph,IIS}(E)$ (**Fig. 7b**). To extract the relevant scattering times, we compare the electrical conductivity computed within the energy/momentum dependent scattering case to the electrical conductivity calculated under the constant relaxation time approximation. In the latter case, the conductivity is directly proportional to the any chosen relaxation time, so when we divide that time out, we are left with an effective constant time that can map to the energy-dependent calculation as:

$$\tau^* = \sigma^{scatt} / (\sigma/\tau_C)^{const} \tag{6}$$

where $\sigma^{scatt}$ is the calculated conductivity including all the carrier energy and momentum dependencies, including either electron-phonon scattering or adding the ionized impurity scattering, while $\sigma^{const}$ is the electrical conductivity from the constant relaxation time approximation calculation. Thus, if the choice is that of a constant relaxation time approximation, that should be doping level (Fermi level) and temperature dependent. In **Fig. 7** the position of the PF peaks is shown by dashed lines, green for the energy dependent scattering time ($\eta_F = 0$ eV) and red for the constant time approximation ($\eta_F = -0.1$ eV). Indicative room temperature relaxation times from the experimental cases of ZrCoBi and ZrCoSb as in **Fig. 3c** are represented by the star-symbols in **Fig. 7**. To plot these, we have scaled the relaxation times in each sub-figure by the relative difference that the power factors had in **Fig. 3c**. Interestingly, the relaxation times for all materials vary according to the Fermi level position from 10s fs up to 100s fs. The shapes of all lines are similar, which might be able to give us sufficient 'statistics' to make crude generalizations. We show that different constant relaxation times values are appropriate for different carrier densities. At low carrier densities (positive $\eta_F$), for example, the suggestion is to employ $\tau_c$ around 200 fs or 100 fs according to the considered scattering physics (phonons only or phonons and impurities). When $E_F = E_C$, we suggest relaxation times around 150 fs if consider the carrier scattering with phonons or in the range 20 to 40 fs if we aim to include the effect of the ionized dopants. This is also where we find that the PF maximum will be encountered. At higher carrier densities, the times drop to 50 fs, and in the case where IIS dominates (which is more relevant for experiments), the effective $\tau^*$ is around 10 to 20 fs. Experimental results also point towards a $\tau^*$ in the 15 to 25 fs range for the experimental doping levels, however, the important point that we can observe here, is that the optimal operation is achieved at different Fermi levels (closer to the green, rather than the red vertical lines).

Finally, it is useful to comment about when the use of the constant relaxation time approximation is adequate, even if the better estimated values from **Fig. 7** are employed. About the effect of temperature, we show in **Fig. 5** that if the relaxation times are scaled by the temperature as $\tau_c(T) = \tau_{c(300)} \times (300/T)$, one can still employ the constant relaxation time approximation, and capture sufficiently qualitatively (with some quantitative features as well) the trends with the carrier density. On the other hand, in the case of a material with a specific carrier density (as in experimental cases) as depicted in **Fig. 6**, the temperature behaviour trends of the power factor are captured by the constant relaxation time approximation at high carrier densities, with the T-scaled rates underestimating the trends and the non-scaled, overestimating them – thus the right trend is somewhere between the two limits. At low carrier densities, as indicated in **Fig. 6b**, the energy-dependent scattering times indicate a downward trend with T, which is not captured by the constant scattering time - in that case a T-scaled rate can be a better choice. Notably one of the best p-type TE half-Heusler alloy, the Ti doped NbFeSb, at its optimal doping ($8 \times 10^{20}$ cm$^{-3}$) and crystalline quality (larger grains), exhibits a PF that decreases with T,[45] in contrast to what the constant relaxation time approximation suggests.

## 4. Conclusions

Using a fully numerical simulator for the extraction of the scattering rates in complex bandstructure materials, we performed a comparison of the thermoelectric power factor of five

Co-based half-Heusler alloys in computational cases which take account the energy/momentum dependences of the scattering times, versus the commonly employed approach of the constant scattering time. We show that the use of constant relaxation scattering time smoothens out the richness of the bandstructure complexities, in addition to lumping all scattering physics (deformation potentials, electron-phonon interactions, ionized impurity scattering, dielectric constants, etc.) into one arbitrary parameter. In this way, it reduces the possibility in employing the bandstructure richness as a design tool, whereas this is more feasible when all energetics of scattering are considered. As a result, when comparing the power factor outcomes, we detect different rankings between the different materials with respect to the power factor maximum, different densities at which that power factor peak is observed, and different temperature trends. We reckon that these evaluations are however sensitive to the accuracy of the inputted deformation potentials. Our analysis emphasizes the relevance of considering the details of the scattering physics by means of a full band energy/momentum dependence of the carriers' scattering time when predicting thermoelectric material properties and would be helpful especially in the identification of materials descriptors for materials screening.

**Supplementary material**: additional information on the computational scheme validation is provided in supplemental online materials.

# Acknowledgements

This work has received funding from the Marie Skłodowska-Curie Actions under the Grant agreement ID: 788465 (GENESIS - Generic semiclassical transport simulator for new generation thermoelectric materials) and from the European Research Council (ERC) under the European Union's Horizon 2020 Research and Innovation Programme (Grant Agreement No. 678763).

| Compound | $D_{ADP}$ [eV] | $v_s$ [m/s] × $10^3$ | $\rho$ [g/cm$^3$] | $D_{ODP}$ [eV/m] × $10^{10}$ | $\omega$ [eV] | $\varepsilon_r$ |
|---|---|---|---|---|---|---|
| NbCoSn | 0.5 | 5.36 | 8.43 | 2.15 | 0.034 | 22.66 |
| ZrCoBi | 0.8 | 3.21 | 9.83 | 1.80 | 0.028 | 20.37 |
| TiCoSb | 0.5 | 4.04 | 7.42 | 2.20 | 0.036 | 19.09 |
| ZrCoSb | 1.0 | 5.55 | 7.14 | 2.05 | 0.028 | 17.87 |
| HfCoSb | 0.4 | 5.64 | 9.54 | 1.85 | 0.028 | 17.51 |

Table I: Material parameters used in the present work (extracted from literature). [26, 27]

Figure 1

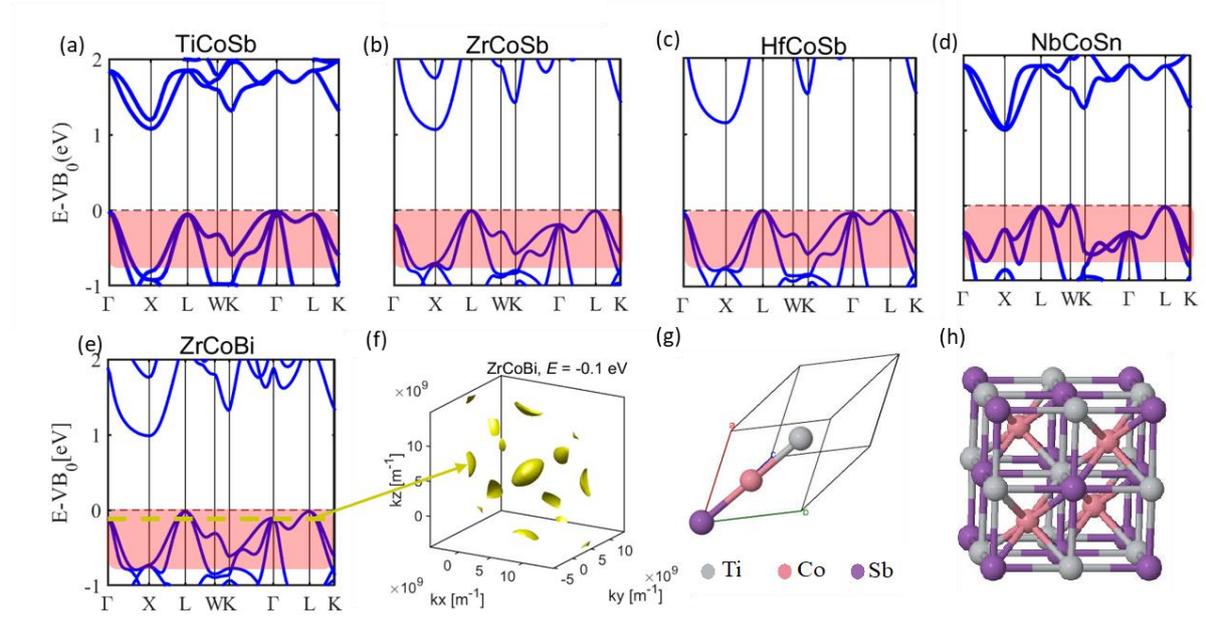

Figure 1 caption:

Bandstructures of (a) TiCoSb, (b) ZrCoSb, (c) HfCoSb, (d) NbCoSn and (e) ZrCoBi. The red-shaded areas that extend up to $E = -0.7$ eV below the valence band edge indicate the energy window used for the transport properties computation. (f) Iso-energy surface of one of the valence band of ZrCoBi at $E = -0.1$ eV below valence band edge represented on the reciprocal unit cell. The multi-valley nature of the bandstructure is clearly evident. (g) Primitive unit cell used for the DFT bandstructure calculations and (h) conventional zincblende unit cell of the TiCoSb. The oblique geometry typical of the zincblende is visible.

Figure 2

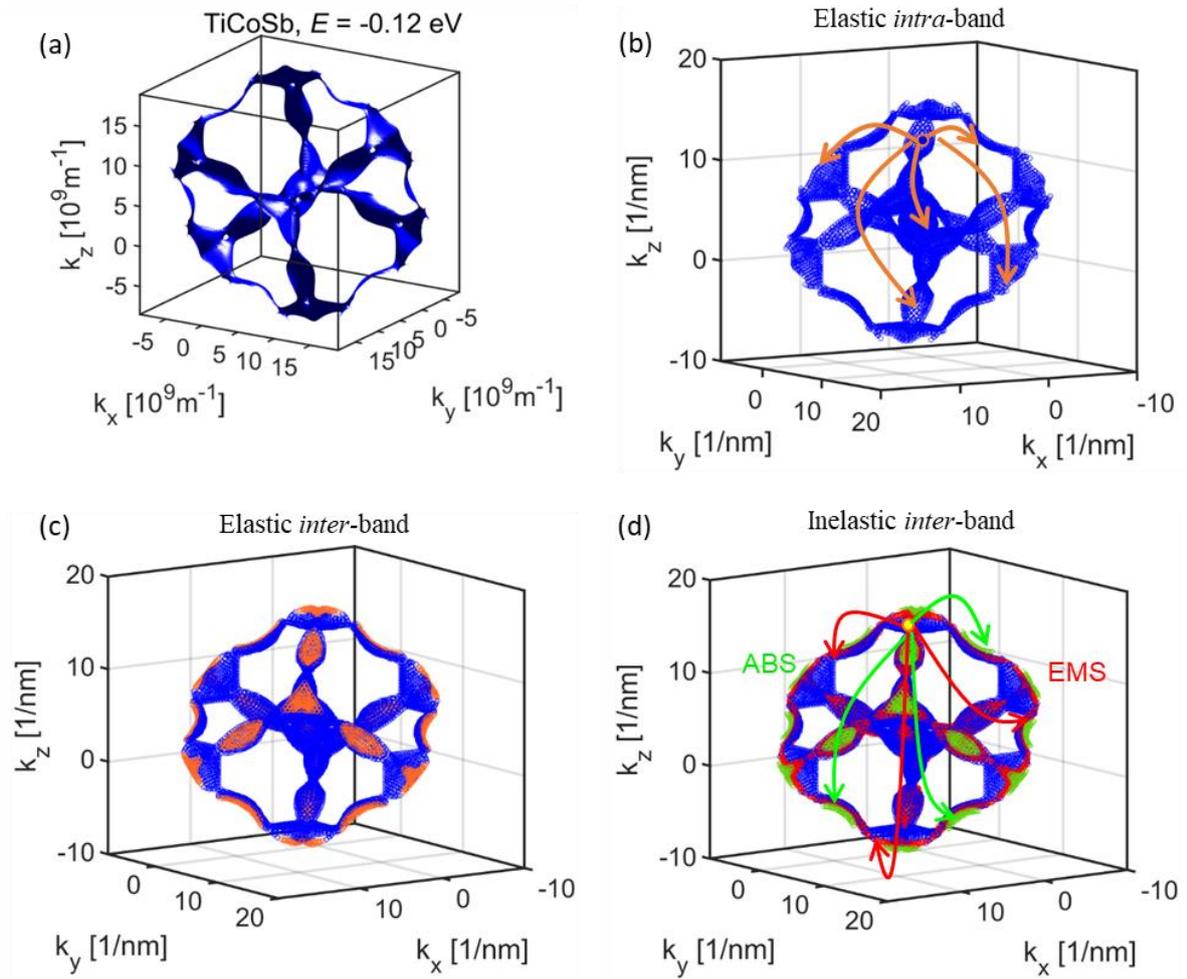

Figure 2 caption:

(a) Warped and highly anisotropic *k*-space iso-energy surface of one of the three valence bands of TiCoSb at $E$ = -0.12 eV below the valence band edge. The surface is formed by interconnected elongated 'tubes' and occupies the entire reciprocal unit cell. (b) The same iso-energy surface as in (a), formed by 'points' indicating the *k*-states that form it. Each dot is a transport state, from which charge carriers (holes) can scatter into other states under an elastic, intra-band scattering event (indicated by the arrows). (c) The same iso-energetic surface as in (a) and (b) in blue, and a different iso-surface at the same energy but belonging to another valence band, shown in orange. Each state (dot) in the blue iso-energy surface can be an initial scattering state and each dot in the orange surface can be a final scattering state during an inter-valley elastic scattering event. (d) The same iso-energy surface in blue, together with two other iso-energy surfaces (green/red) from another valence band and at different energies.

Transitions from the blue dots into the green/red dots can be the usual transitions in the case of an inelastic inter-band scattering event (i.e. absorption/emission of optical phonons).

Figure 3

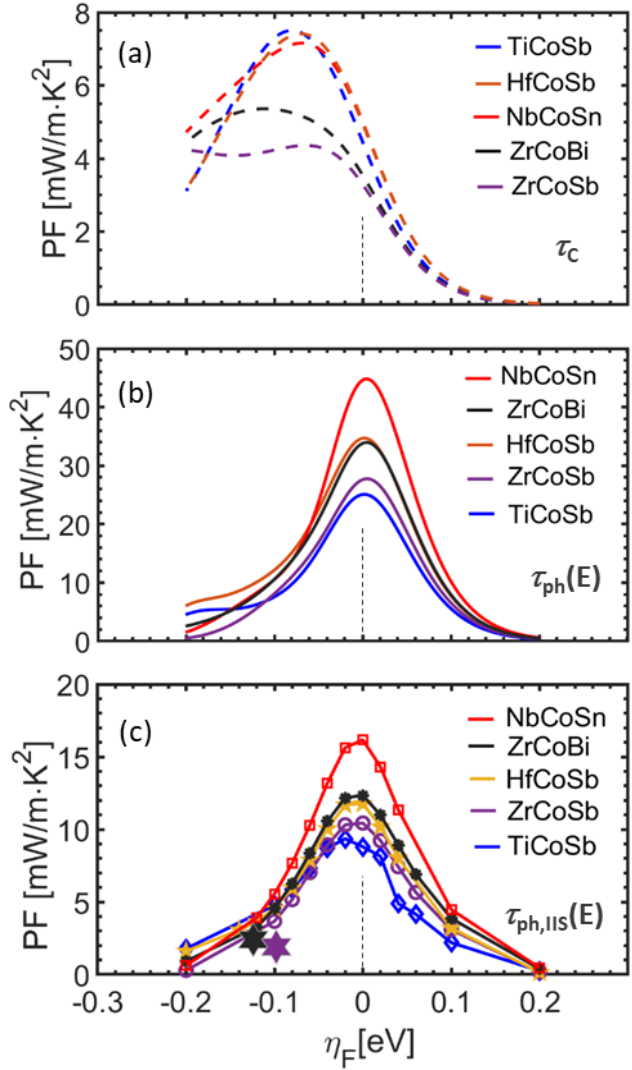

Figure 3 caption:

Power Factor PF versus relative position of the Fermi level, $\eta_F$, for three different scattering scenarios. $\eta_F = 0$ means that the Fermi level is at the band edge – indicated by the vertical dashed lines. Positive $\eta_F$ values mean that the Fermi level is into the gap while a negative $\eta_F$ means the Fermi level is into the valence band. Results for TiCoSb are shown in blue, for HfCoSb in orange, for NbCoSn in red, for ZrCoBi in black, and for ZrCoSb in purple. (a) The constant relaxation time approximation ($\tau_c$ = 10 fs) case. (b) The energy dependent electron-phonon scattering case. (c) The ionized impurity scattering is added to electron-phonon scattering of (b). Experimental PF values at 300 K reported for ZrCoBi and ZrCoSb are shown by the black and purple hexagons,[35] respectively. For those, the $\eta_F$ value corresponds to the measured carrier density: $p = 2.20 \times 10^{21}$ cm$^{-3}$, $\eta_F$ = -0.124 eV for ZrCoBi, $p = 1.47 \times 10^{21}$ cm$^{-3}$, $\eta_F$ = -0.098 eV for ZrCoSb, for a stoichiometric Sn doping of 0.15.[35]

Figure 4

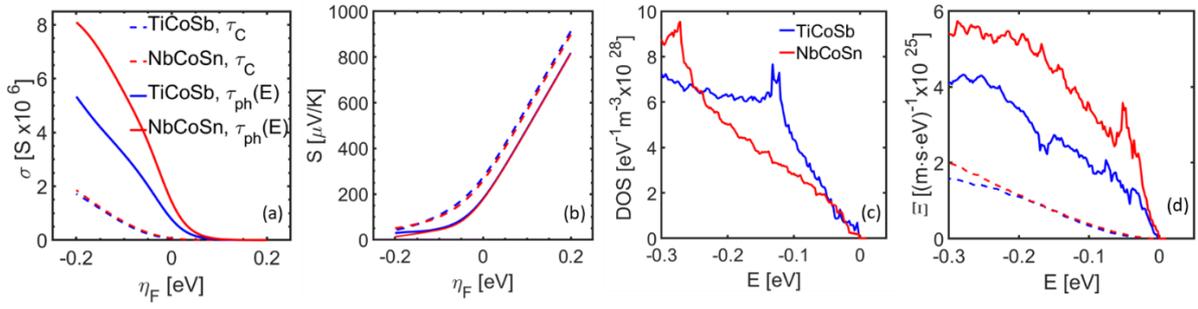

Figure 4 caption:

(a) Electrical conductivity $\sigma$, (b) Seebeck coefficient $S$, (c) density-of-states DOS, and (d) transport distribution functions $\Xi(E)$ for NbCoSn in red and TiCoSb in blue. Solid lines are for the phonon-limited energy-dependent relaxation time calculations ($\tau_{ph}(E)$), while dashed lines are for the constant relaxation time approximation ($\tau_c$).

Figure 5

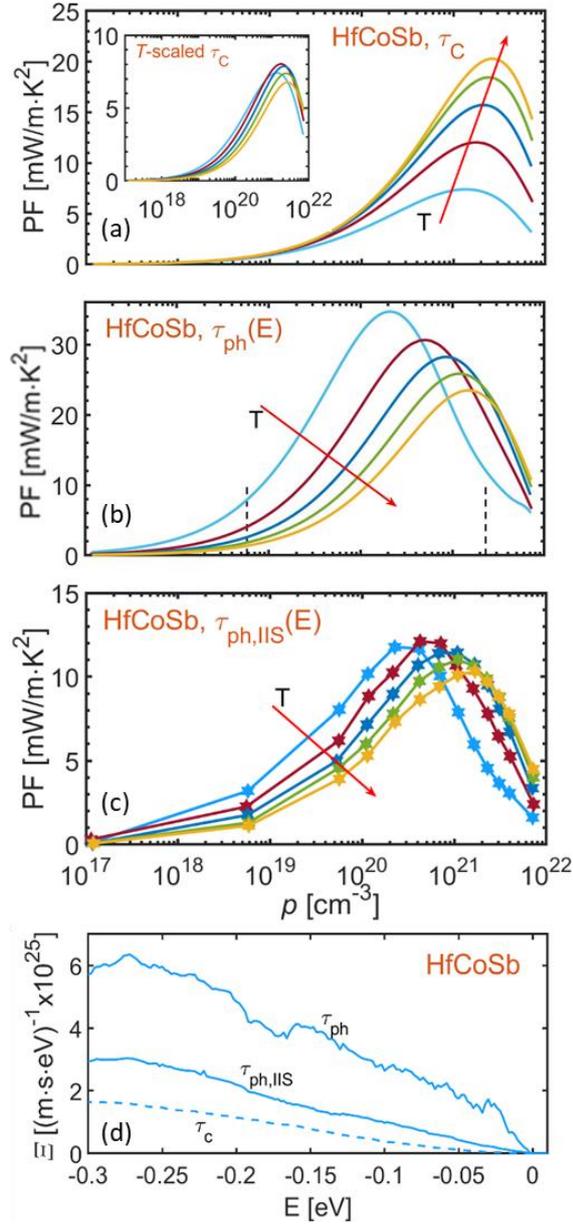

Figure 5 caption:

Power factor PF for HfCoSb versus carrier concentration ($p$) plotted at different temperatures for different scattering scenarios: (a) the constant relaxation time approximation, (b) the energy-dependent electron-phonon scattering $\tau_{ph}$, (c) the electron-phonon scattering and the ionized impurities scattering $\tau_{ph,IIS}$. The red arrow indicates the direction of increasing temperature $T$ from 300 K to 900 K. Inset of (a): the case of constant relaxation time when the scattering time is linearly scaled by the temperature. (d) Transport distribution functions $\Xi(E)$ for the scattering cases (solid line) and the constant scattering time (dashed line) at 300 K. In the case with the IIS, the plotted data are for the Fermi level at the band edge, $p = 2.3\times10^{20}$ cm$^{-3}$.

Figure 6

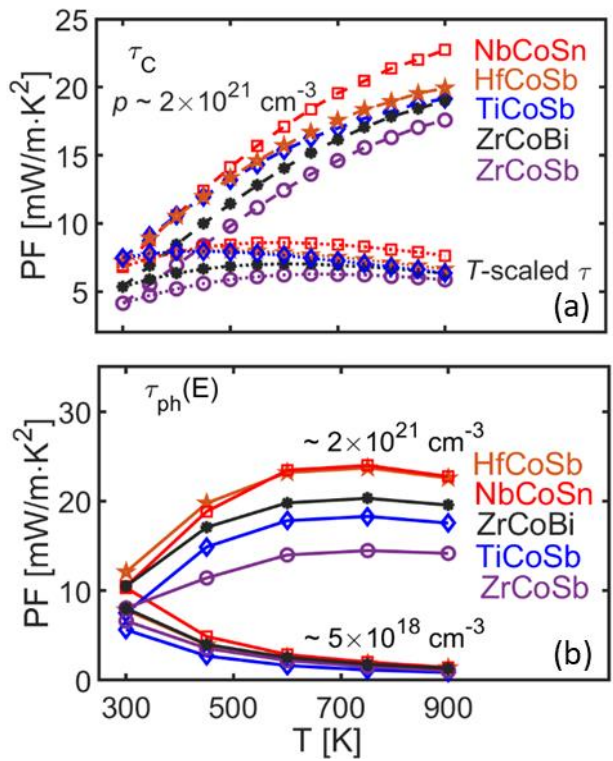

Figure 6 caption:

Power factor PF versus temperature at constant carrier density for two scattering scenarios: (a) Constant relaxation time ($\tau_c$ = 10 fs, dashed lines) and with its linear temperature scaling versus the temperature (dotted lines). (b) Case for electron-phonon scattering, for two different doping concentrations corresponding to non-degenerate and degenerate conditions, indicated by the dashed lines in **Fig. 5b**.

Figure 7

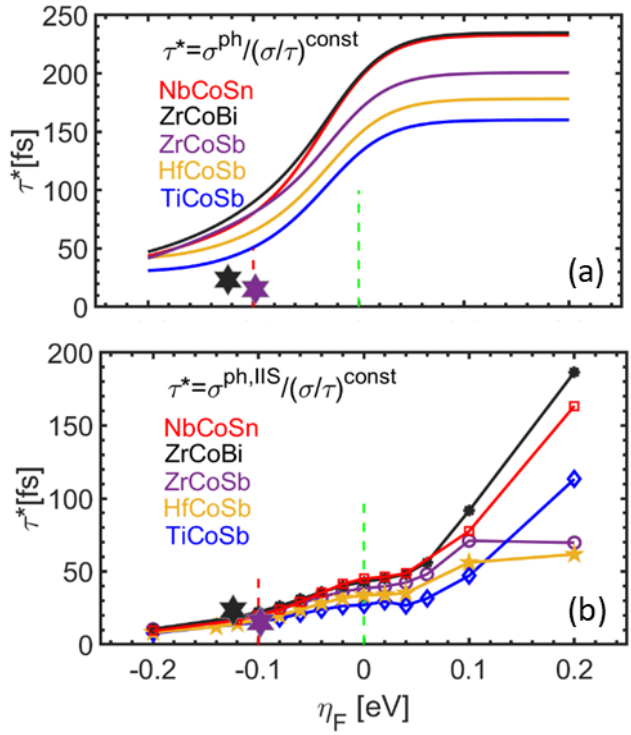

Figure 7 caption:

Equivalent Fermi level dependent relaxation times extracted from Eq. (6) at 300 K. (a) The equivalent constant times computed by considering only phonon scattering. (b) The equivalent times computed by considering both phonon and impurity scattering. The green and red dashed lines correspond to the peak position of the PF under the energy-dependent relaxation time calculations, and the constant relaxation time calculations, respectively (see **Fig. 3**). The stars represent the relaxation time obtained for the experimental cases depicted in the inset of **Fig. 3c**, with black for ZrCoBi and purple for ZrCoSb.